\documentclass[10pt, paper=a4, UKenglish]{article}
\usepackage{graphicx}
\def\Title#1{\begin{center} {\Large #1 } \end{center}}
\def\Author#1{\begin{center}{ \sc #1} \end{center}}
\def\Address#1{\begin{center}{ \it #1} \end{center}}

\newcommand\pubblock{\rightline{\begin{tabular}{l} Proceedings of the CTD/WIT 2019\\ \pubnumberctd \\ \pubnumbercms \\
         \pubdate  \end{tabular}}}

\newenvironment{Abstract}{\begin{quotation} \begin{center} 
             \large ABSTRACT \end{center}\bigskip 
      \begin{center}\begin{large}}{\end{large}\end{center} \end{quotation}}

\newenvironment{Presented}{\begin{quotation} \begin{center} 
             PRESENTED AT\end{center}\bigskip 
      \begin{center}\begin{large}}{\end{large}\end{center} \end{quotation}}

\def\beq{\begin{equation}}
\def\eeq#1{\label{#1}\end{equation}}
\def\eeqn{\end{equation}}

\def\beqa{\begin{eqnarray}}
\def\eeqa#1{\label{#1}\end{eqnarray}}
\def\eeqan{\end{eqnarray}}

\let\bar=\overbar

\def\Dslash{\not{\hbox{\kern-4pt $D$}}}
\def\dslash{\not{\hbox{\kern-2pt $\del$}}}

\def\msb{{\bar{\ssstyle M \kern -1pt S}}}

\usepackage{color}
\usepackage{lineno}
\usepackage{subfig}
\usepackage{hyperref}

\newcommand\pubnumberctd{PROC-CTD19-088}
\newcommand\pubnumbercms{CMS CR-2019/041}

\newcommand\pubdate{\today}

\def\affiliation{
On behalf of the CMS Collaboration, \\
Dipartimento di Fisica \\
Universit\`a di Torino and INFN Sezione di Torino, Italy}

\newcommand{\conference}{Connecting the Dots and Workshop on Intelligent Trackers (CTD/WIT 2019)\\
Instituto de F\'isica Corpuscular (IFIC), Valencia, Spain\\ 
April 2-5, 2019}

\usepackage{fancyhdr}
\definecolor{mygrey}{RGB}{105,105,105}
\begin{document}


\large
\begin{titlepage}
\pubblock

\vfill
\Title{The CMS Tracker Upgrade for the High-Luminosity LHC}
\vfill

\Author{Ernesto Migliore}
\Address{\affiliation}
\vfill

\begin{Abstract}
The LHC machine is planning an upgrade program, which will smoothly
bring the instantaneous luminosity to about $5-7.5\times10^{34}~\mathrm{cm}^{-2}\mathrm{s}^{-1}$ in 2028,
to possibly reach an integrated luminosity of 3000-4500 fb$^{-1}$ by
the end of 2039. This High-Luminosity LHC scenario, HL-LHC, will
require a preparation program of the LHC detectors known as Phase-2
upgrade. The current CMS Outer Tracker, already running beyond design
specifications, and the recently installed CMS Phase-1 Pixel Detector will
not be able to survive the HL-LHC radiation conditions. Thus, CMS will
need completely new devices in order to fully exploit the
high-demanding operating conditions and the delivered luminosity. The
new Outer Tracker should also have trigger capabilities. To achieve
such goals, the R\&D activities have investigated different options for
the Outer Tracker and for the pixel Inner Tracker. The developed
solutions will allow including tracking information at the Level-1
trigger. The design choices for the Tracker upgrades are discussed
along with some highlights on the technological choices and the R\&D
activities. 
\end{Abstract}

\vfill

\begin{Presented}
\conference
\end{Presented}
\vfill
\end{titlepage}
\def\thefootnote{\fnsymbol{footnote}}
\setcounter{footnote}{0}
%

\normalsize 


\section{Introduction}
The goal of the High-Luminosity LHC program is to collect an
integrated luminosity of 3000~fb$^{-1}$, optionally up to 4500~fb$^{-1}$, in about
ten years of operations starting in 2028 and with a 
peak luminosity of $7.5 \times 10^{34}~\mathrm{cm}^{-2}\mathrm{s}^{-1}$.
As the bunch crossing separation will stay the same as today (25 ns),
the increase of instantaneous luminosity will result in up to 200 inelastic proton-proton collisions per bunch crossing (pileup)
while the integrated luminosity will lead to an unprecedented hostile
radiation environment.
In the most exposed point in the Tracker
volume, at $r \simeq 3$ cm from the beamline, the expected fluence
after 3000~fb$^{-1}$ will reach $2.3 \times 10^{16}~\mathrm{1~MeV~n_{eq}/cm^{2}}$ 
and the corresponding total ionizing dose (TID) 12~MGy. 

\begin{figure}[!htb]
  \centering
  \includegraphics[width=0.75\linewidth]{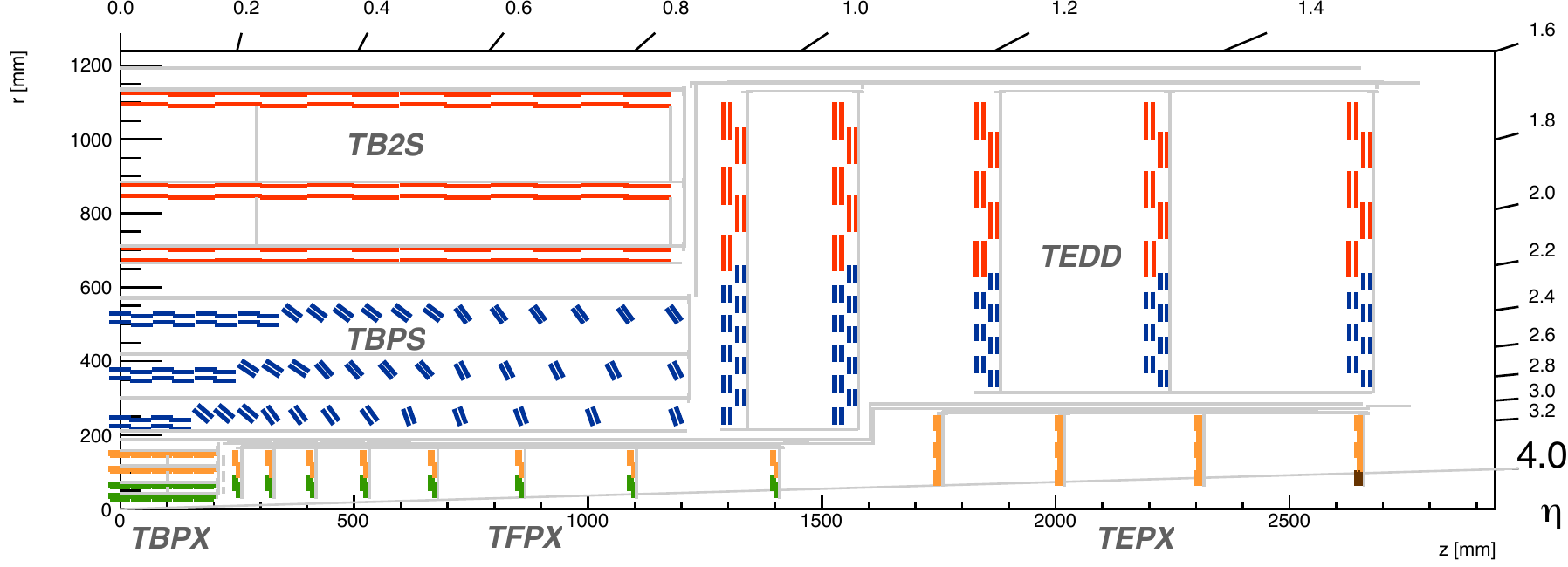}
  \caption{Sketch of one quarter of the layout of the CMS Tracker for
    HL-LHC in the $r-z$ view.
    Inner Tracker 1x2 and 2x2 readout chips modules are shown in green
    and yellow respectively, Outer Tracker PS and 2S modules in blue
    and red. Details of the Inner Tracker and Outer Tracker modules
    can be found in the main text.}
  \label{fig:TkLayout}
\end{figure}

The proposed layout of the Tracker~\cite{CMS:TP,CMS:TDR-014} is shown in Fig.\ref{fig:TkLayout}: the
Inner Tracker (IT) region, $r<$20~cm ($r<$30~cm for $|z|>$120~cm), will be instrumented with 
high granularity pixel detectors guaranteeing an efficient pattern
recognition in the high track density environment of 
HL-LHC; the Outer Tracker (OT) region will be instrumented with $p_T$-modules
providing a ${\cal O}(10)$ on-detector data reduction so that the precise
measurements of the trajectories of charged particles from the Tracker
can be used in the Level-1 (L1) trigger of the CMS experiment.
Specifically, the main features of the upgraded Tracker will be:
\begin{itemize}
\item contribute to L1 trigger measuring at 40 MHz the momenta of charged
  particles and rejecting those with $p_T<$2 GeV/$c$; 
\item be compatible with the upgraded L1 trigger of CMS which
  requires large readout bandwidth to withstand a 750 kHz L1 accept rate
  (currently about 100 kHz), and deep front-end buffers to comply with
  an increased L1 latency of 12.5 $\mu$s (currently 4 $\mu$s);
\item ensure a robust two-track separation especially in high energy jets by
  means of a high granularity/low occupancy detector, ${\cal O}(0.1\%)$ in the Inner
  Tracker and ${\cal O}(1\%)$ in the Outer Tracker respectively;
\item mitigate efficiently the pileup up to $|\eta|$=4;
\item guarantee an accurate measurement of the momentum and maintain a
  low level of fake tracks through optimal layout and reduced
  material budget.
\end{itemize}

The key ingredients to guarantee the functionality of the Tracker during the entire HL-LHC lifetime are:
\begin{itemize}
\item the deployment of dedicated sensors and ASICs;
\item the operations of the sensors at the temperature of ~-20~$^\circ$C;
\item a design which allows for full access to the Inner Tracker for
  maintenance and, optionally, for the replacement of the innermost
  layer of the barrel after half the HL-LHC timespan.
\end{itemize}


\section{The Inner Tracker}
The Inner Tracker is the main detector used in the offline track
reconstruction for the pattern recognition. The IT will be made of hybrid pixel modules
comprising a pixel sensor and two (1x2) or four (2x2) readout chips.
The IT is structured into a barrel
(TBPX) consisting of four layers with 9 modules/ladder and no projective gap
at $|\eta|$=0, a forward (TFPX) made of 8(x2) small discs with 4
rings/disc, and a luminosity extension (TEPX) made of 4(x2) large discs
with 5 rings/disc. The innermost ring of the last TEPX disc will be
entirely devoted to the measurement of the bunch-by-bunch luminosity.
Overall, the IT will cover a surface of about 4.9~$\mathrm{m}^2$ for a total of
2B channels.
\subsection{Inner Tracker sensors and readout chip}
\begin{figure}[!htb]
  \centering
  \subfloat[]{\includegraphics[height=0.33\linewidth]{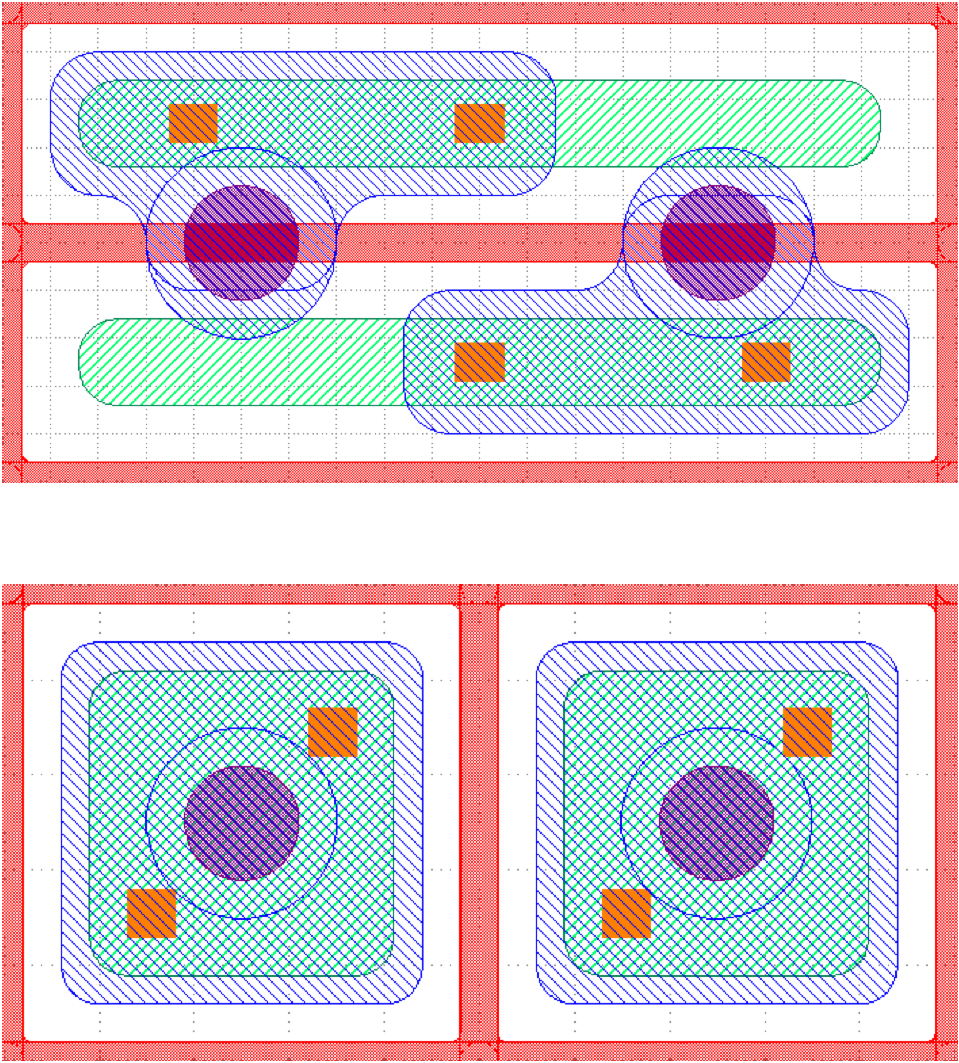}}
  \qquad \qquad
  \subfloat[]{\includegraphics[height=0.34\linewidth]{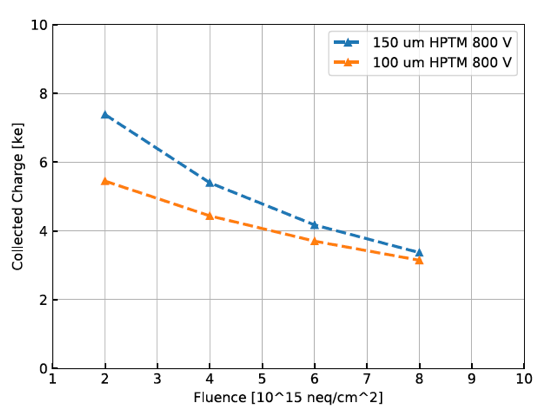}}
  \caption{IT sensors: layout of the 25x100~$\mu$m$^2$ and
    50x50~$\mu$m$^2$ pixel cells (a); simulated collected charge
    versus fluence in planar n-on-p pixel sensors biased at 800~V
    predicted by the \textsl{Hamburg Pentatrap Model}~\cite{IEEE2018:JSchwandt} (b).} 
  \label{fig:ITsensors}
\end{figure}

\paragraph{IT sensors.} High granularity will be achieved using small
pitch pixel cells with either a rectangular (25$\times$100 $\mu$m$^2$) or a square
(50$\times$50 $\mu$m$^2$) aspect ratio, both compatible with the same
bond pattern (Fig.~\ref{fig:ITsensors} (a)).
The baseline choice for the sensor technology is to adopt planar n-on-p silicon sensors with 100-150~$\mu$m active thickness.
Results from TCAD simulations shown in Fig.~\ref{fig:ITsensors} (b) indicate that charge
above three times the foreseen threshold of the front-end chip can be
collected up to a fluence of $0.8 \times 10^{16}~\mathrm{1~MeV~n_{eq}/cm^{2}}$,
that is approximately the fluence expected in the most exposed regions of the
IT after half the HL-LHC timespan, operating the detector at 800~V.
Since the sensors will be produced from single-side processing, no guard-rings on
the backplane will limit the electric potential on the cut edge.
Therefore protection mechanisms to avoid sparks between the readout
chip, at ground potential, and the sensor are under investigation.
For the innermost modules, usage of 3D silicon sensors, which allow efficient charge collection
after irradiation operating at 150-200~V, is under consideration.

\paragraph{IT readout chip.} The readout chip (ROC) is required to
operate with low noise at low thresholds (1000~e) for efficient
detection of signal on irradiated sensors, to sustain high hit rate
(99\% efficiency at 3.2 GHz/cm$^2$) and to have high trigger rate capabilities
(thanks to four 1.28 Gb/s output links).
With the option of replacing the innermost layer of the TBPX,
radiation tolerance requirements can be halved with respect to the 12~MGy
TID expected for 3000 fb$^{-1}$.
The ROC is currently being developed in CMOS 65~nm technology by the
RD53/CERN collaboration~\cite{RD53:proposal,RD53:RD53Aspec}. In 2018 a first prototype (RD53A) with
dimensions 1x2~cm$^2$, half-size of the final chip,
has been delivered for standalone qualification and for use in
single-chip assemblies with IT prototype sensors. All the three
front-end architectures implemented in RD53A resulted fully functional
after a TID of 5~MGy. 
Measurements with X-rays show that the RD53A chip meets the
required specifications on the hit rate~\cite{Pixel2018:DRuini}. Single chip assemblies with 3D sensors irradiated up to $1 \times 10^{16}~\mathrm{1~MeV~n_{eq}/cm^{2}}$ 
and exposed to 120 GeV/$c$ proton beam at CERN SPS indicate that
post-irradiation the decrease in the hit efficiency can be as low as 1\%
when operating the sensor at a moderate bias voltage of 150~V~\cite{VCI2019:JCampderros}.

\subsection{Inner Tracker modules}
\begin{figure}[!htb]
  \centering
  \includegraphics[angle=270, width=0.7\linewidth]{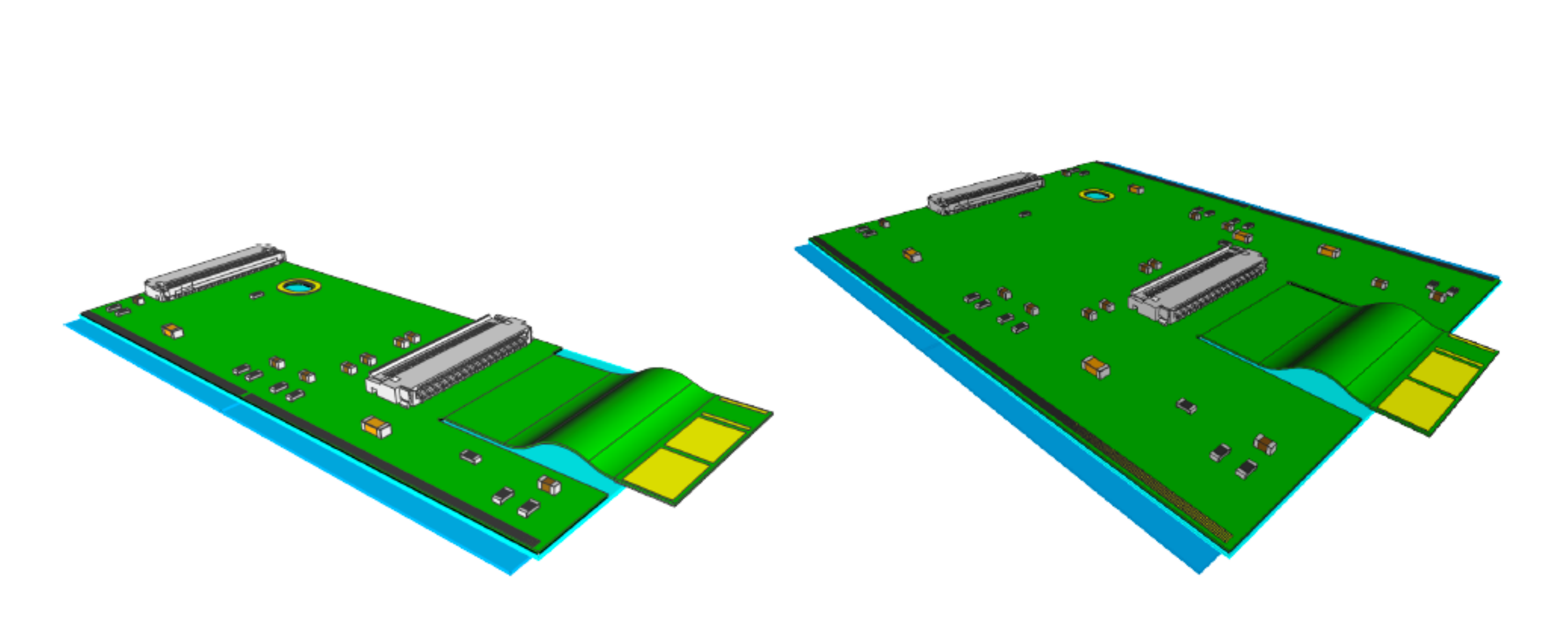}
  \caption{IT modules: rendering of the 1x2 (left) and 2x2 (right)
    readout chips modules.}
  \label{fig:ITmodules}
\end{figure}
Figure~\ref{fig:ITmodules} shows the two types of modules, 1x2 and 2x2 ROCs,
which will equip the IT.
The module consists of the silicon sensor, a High-Density
Interconnect circuit and the readout chip. To reduce the power loss in the cable
with an acceptable cable mass, IT modules will be powered serially in
groups of 8-12 modules each with the ROCs on the same module being powered
in parallel. Since the ROC is the only ASIC present on the module, the
1.2~V required by the front-end is provided by a Low Drop Out (LDO) regulator
integrated in the ROC together with a shunt designed to absorb the
extra current flowing in case of a failure of one of the chips. 
Care in the design of the module is taken to guarantee that
the cooling circuit is located under the region where the Shunt-LDOs,
one for the analog and one for the digital domain, are placed in the
ROC and hotspots in the power dissipation are expected. 

\section{The Outer Tracker}
The Outer Tracker will measure in real-time all charged particles tracks 
with $p_T>$2 GeV/$c$ for transmission to the L1 trigger.
This will enable CMS to maintain low trigger rates, without lowering
the efficiency for the physics of interest, even in the harsh environment expected at the HL-LHC.
In the offline track reconstruction the OT will be the main detector for the
measurement of the momentum of charged particles.
The OT is arranged in a barrel made of two subsystems, TBPS for the innermost 3 layers and
TB2S for the outermost 3 layers, and 5(x2) endcap discs (TEDD).
The OT will cover a surface of about 192~$\mathrm{m}^2$ for a total of
42M strip and 170M macro-pixel channels.
\begin{figure}[!htb]
  \centering
  \includegraphics[width=0.7\linewidth]{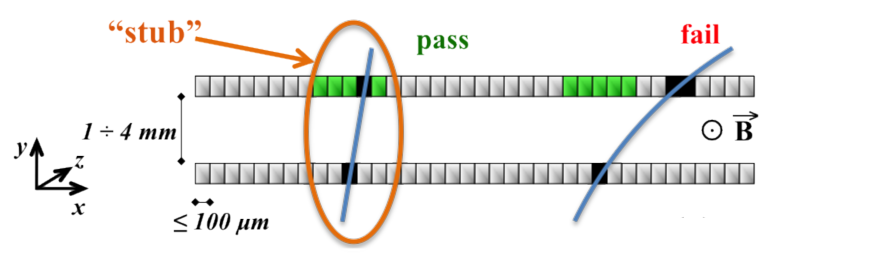}
  \caption{Sketch of a $p_T$-module showing the concept of stub
    selection.}
  \label{fig:pTmodules}
\end{figure}
\subsection{Outer Tracker modules and sensors}
\paragraph{The \boldmath{$p_T$}-modules.} The reduction required to limit the data volume sent out at the
40 MHz collision rate is achieved on the OT modules which are designed
to reject signals from particles below a certain threshold on $p_T$ exploiting the 3.8~T magnetic field of the CMS
solenoid.
A $p_T$-module is made of two silicon sensors with a small spacing in
between: a flex hybrid routes data from both sensors to one ASIC which
looks in a programmable search window for correlated clusters from the
top and the bottom sensor. Only correlated
clusters from high $p_T$ tracks, ``track stubs'', are read out at
40~MHz and sent to the L1 track finder (Fig.~\ref{fig:pTmodules}).
Different sensor spacings, from 1.6~mm up to 4~mm, guarantee a
homogeneous selection of the $p_T$-threshold in different regions 
of the OT. Furthermore, TBPS adopts a tilted barrel geometry to
guarantee full efficiency for stubs from inclined tracks which
otherwise would cross top and bottom sensors in different modules.

\begin{figure}[!htb]
  \centering
  \includegraphics[width=0.4\linewidth]{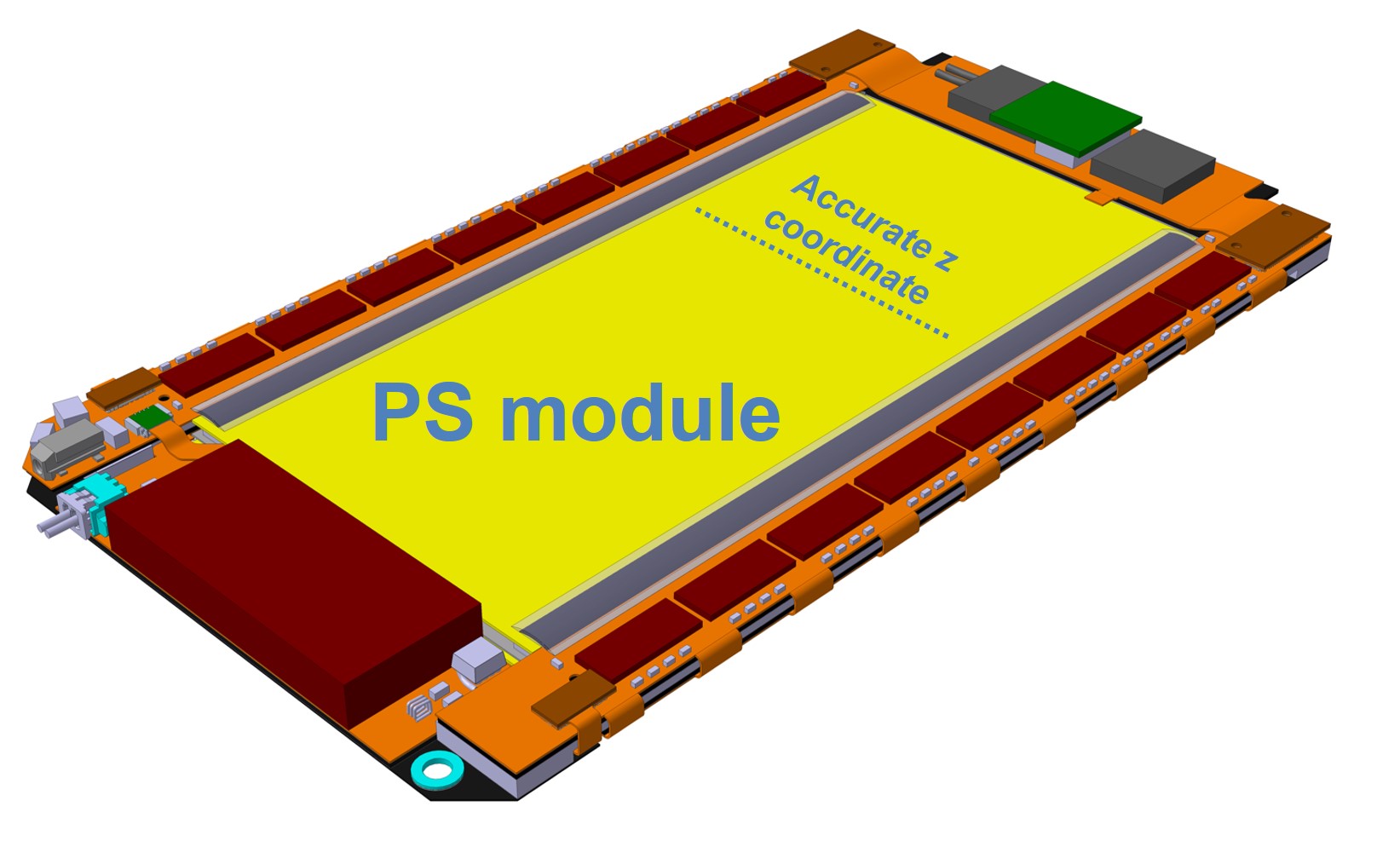}
  \includegraphics[width=0.4\linewidth]{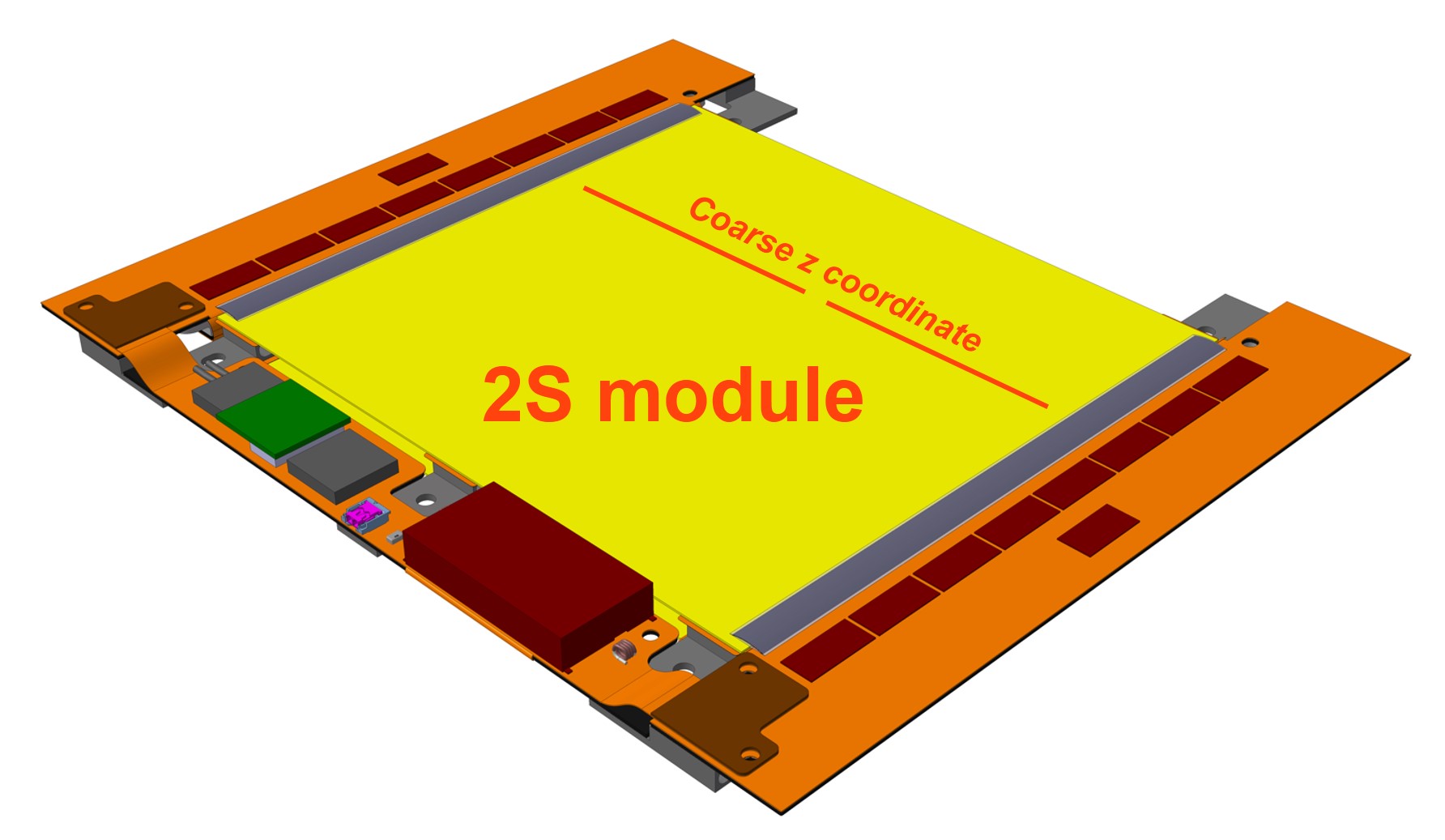}
  \caption{OT modules: rendering of the PS (left) and 2S (right) modules.}
  \label{fig:OTmodules}
\end{figure}

Two module types are foreseen for the OT (Fig.~\ref{fig:OTmodules}):
\begin{itemize}
\item PS modules equipping the TBPS and the innermost TEDD rings.
  A PS module is composed of one strip sensor (two rows of 960
  AC-coupled strips with dimensions 2.35~cm$\times$100~$\mu$m) and a
  one macro-pixel sensor (30 720 DC-coupled pixels with dimensions 1.5~mm$\times$100~$\mu$m).  
\item 2S modules equipping the TB2S and the outermost TEDD rings.
  A 2S module is composed of two strip sensors (two rows of 1016
  AC-coupled strips with dimensions 5~cm$\times$90~$\mu$m per sensor).
\end{itemize}
As no aggregator cards are foreseen between the module and the
back-end, OT modules are self-contained units made of 
the two sensors, the Front-End Hybrids with the readout and concentrator
ASICs, the Service Hybrids hosting the DC-DC power converters and the
data-links, and the spacers providing the proper lever-arm between the
two sensors.
To guarantee the correct functioning of the stub finding procedure, the maximum
rotation allowed between the top and the bottom sensor is 400~$\mu$rad in 2S modules
and 800~$\mu$rad in the shorter PS modules.

\begin{figure}[!htb]
  \centering
  \includegraphics[width=0.4\linewidth]{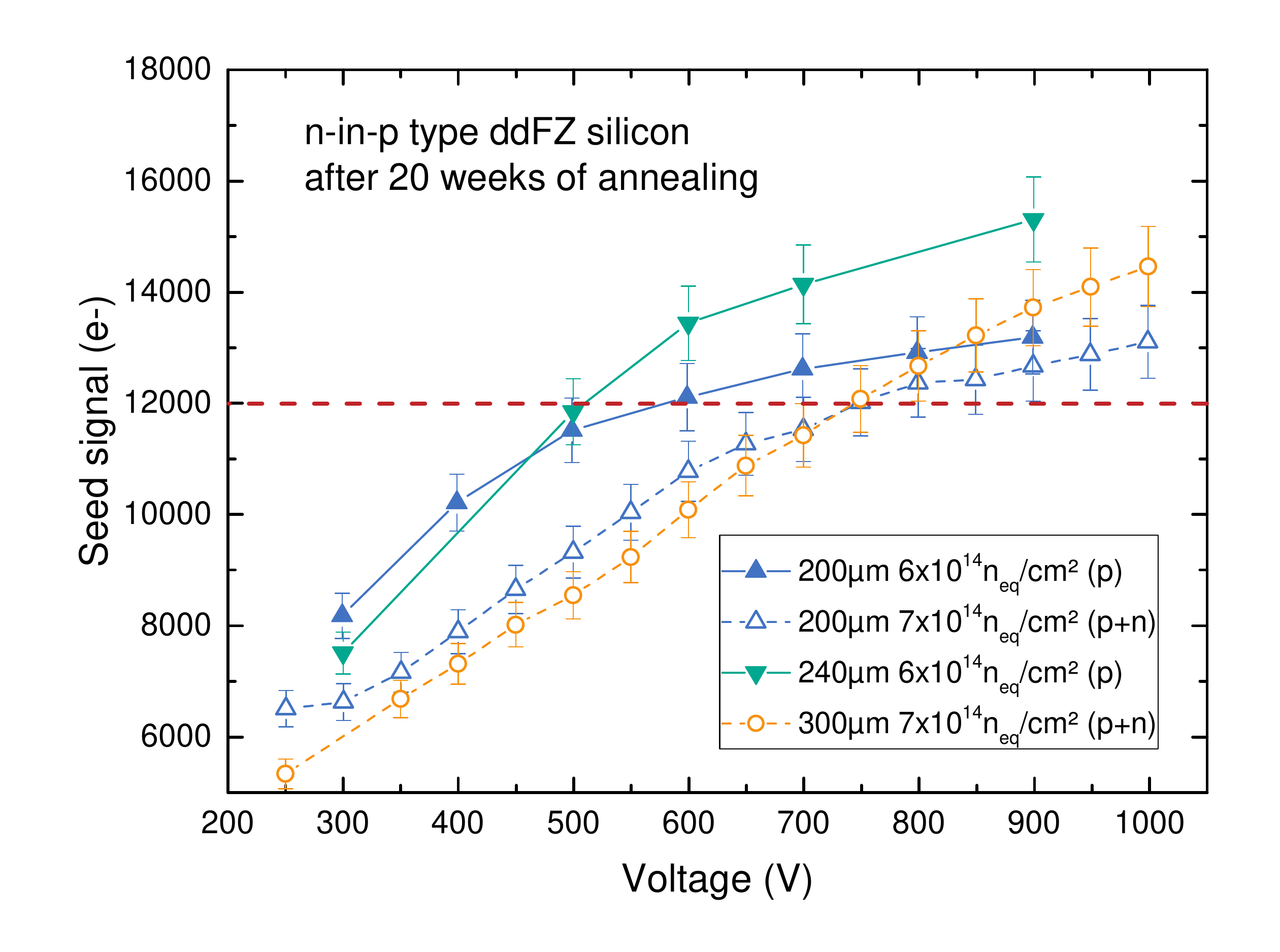}
  \includegraphics[width=0.4\linewidth]{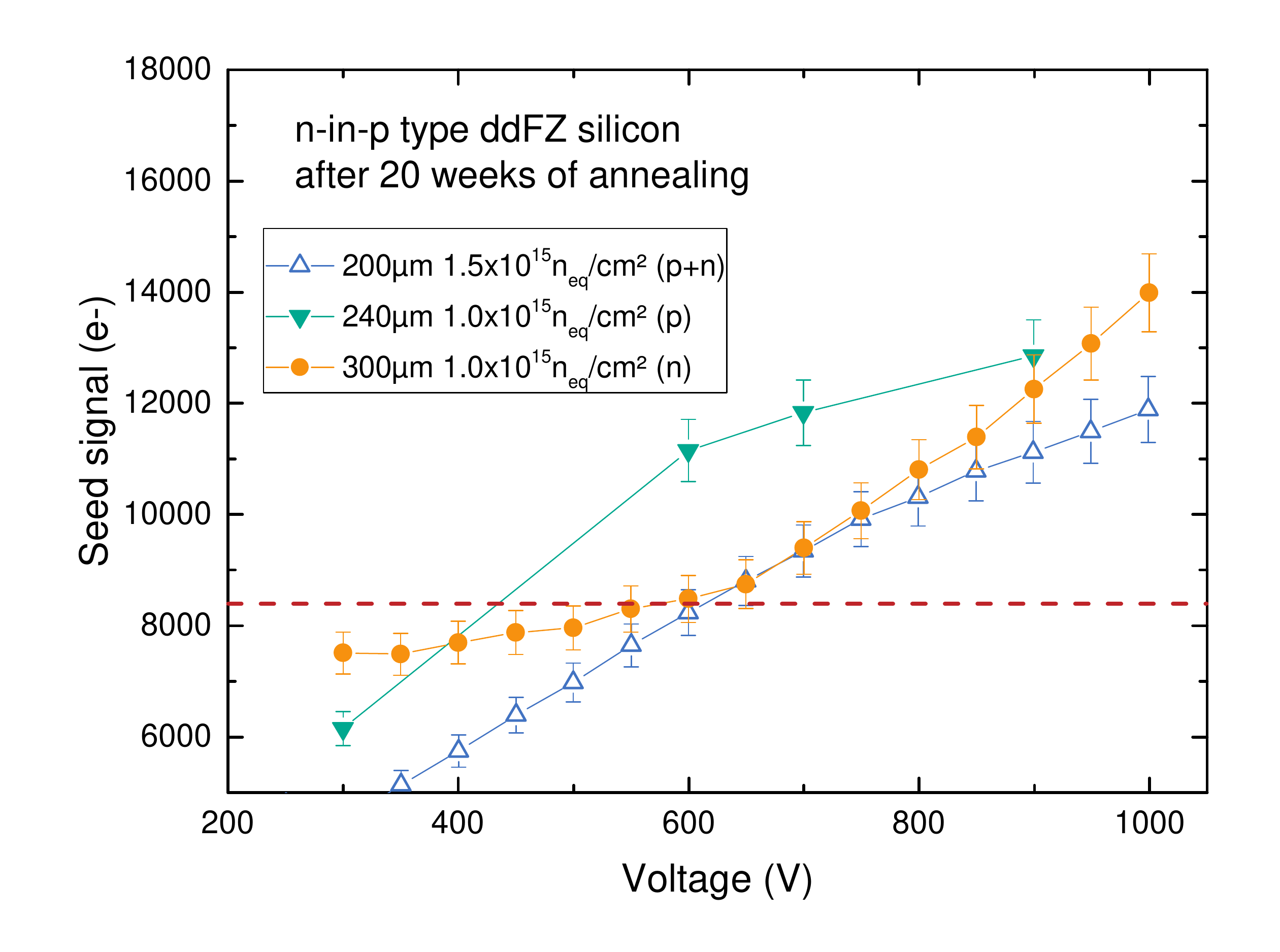}
  \caption{OT sensors: charge collected on the seed strip as a
    function of the bias voltage at fluences relevant for the 2S
    (left) and PS (right) modules respectively. Sensors were
    irradiated up to the nominal fluence with 23 MeV protons (``p'') or with protons plus
    reactor neutrons (``p+n''). The horizontal dashed
    line represents the envisaged seed signal charge (12000~e for 2S modules
    and 8400~e for PS modules respectively).}
  \label{fig:OTsensors}
\end{figure}

\paragraph{The OT sensors.} Since all the front-end ASICs of the OT
adopt a binary readout architecture, the relevant figure-of-merit for
the OT sensors is the charge collected after irradiation on a single
``seed'' channel. This poses more stringent constraints on the strip sensors
used for 2S module and for the PS-strip module which are readout with
ASICs with larger equivalent noise charge (ENC) values (ENC=1000~e in case of the CBC chip used
for the 2S sensors and ENC=700~e for the SSA chip used for the PS-strip sensors).
Figure~\ref{fig:OTsensors} shows the signal on the seed strip as a function of
the bias voltage for n-in-p strip sensor prototypes irradiated at fluence
between 1.5 and 2 times the fluence expected in TBPS and TB2S after
3000~fb$^{-1}$: assuming a threshold 4 times the ENC, sensors with
200-240~$\mu$m active thickness operated at the nominal bias
voltage (600~V) guarantee to collect a charge 3 times larger than
the threshold at the end of the HL-LHC operations.
For the sensors used for the macro-pixels, the front-end ASIC (MPA)
has a much lower noise figure (ENC=200~e). In this case 200~$\mu$m
thick sensors are sufficient to guarantee the requested charge.

\subsection{Outer Tracker back-end electronics}
\begin{figure}[!htb]
  \centering
  \includegraphics[angle=90,width=0.9\linewidth]{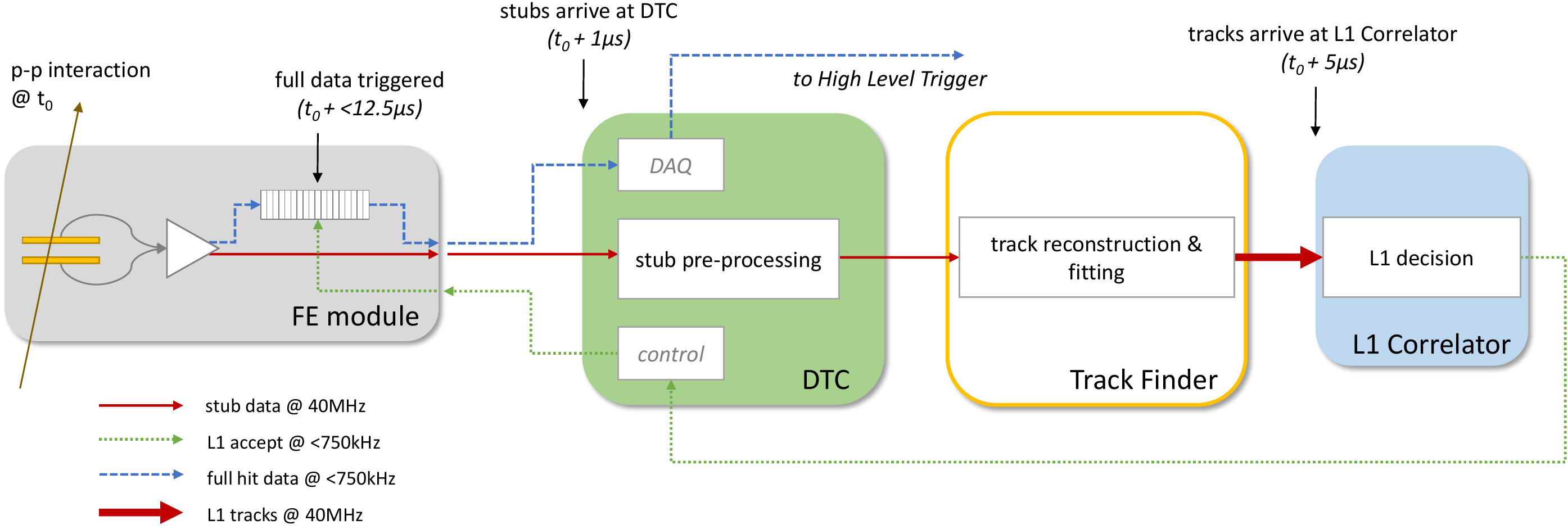}
  \caption{Sketch of the data flow from OT modules to the CMS L1 trigger.}
  \label{fig:OTdataflow}
\end{figure}
A schematic overview of data flow from the Tracker to the CMS L1 trigger is shown in Fig.~\ref{fig:OTdataflow}.
Stubs from up 72 OT modules are sent through optical fibers to one
Data Trigger and Control board (DTC) which unpacks the stubs and converts the information from local to global
coordinates. The DTC forwards the data, sliced in space and in time,
to a Track Finder Processor (TFP).
Each input of the TFP collects the data stream from a single event and from
two overlapping nonants in which the OT is cabled. Pattern
recognition and track finding is performed by the TFP using commercially available
FPGAs. Two different approaches for the pattern recognition and track
fitting have been perfected after the TDR and verified in real life on
two TFP hardware demonstrators. Both the systems have proved to be
fully functional delivering tracks to the L1 trigger system within the
allotted time-budget of 4~$\mu$s. Currently, a hybrid approach
combining the best of both algorithms is under study~\cite{CTD2019:TJames}.

\section{Common aspects and performance}
\begin{figure}[!htb]
  \centering
  \subfloat[]{\includegraphics[width=0.4\linewidth]{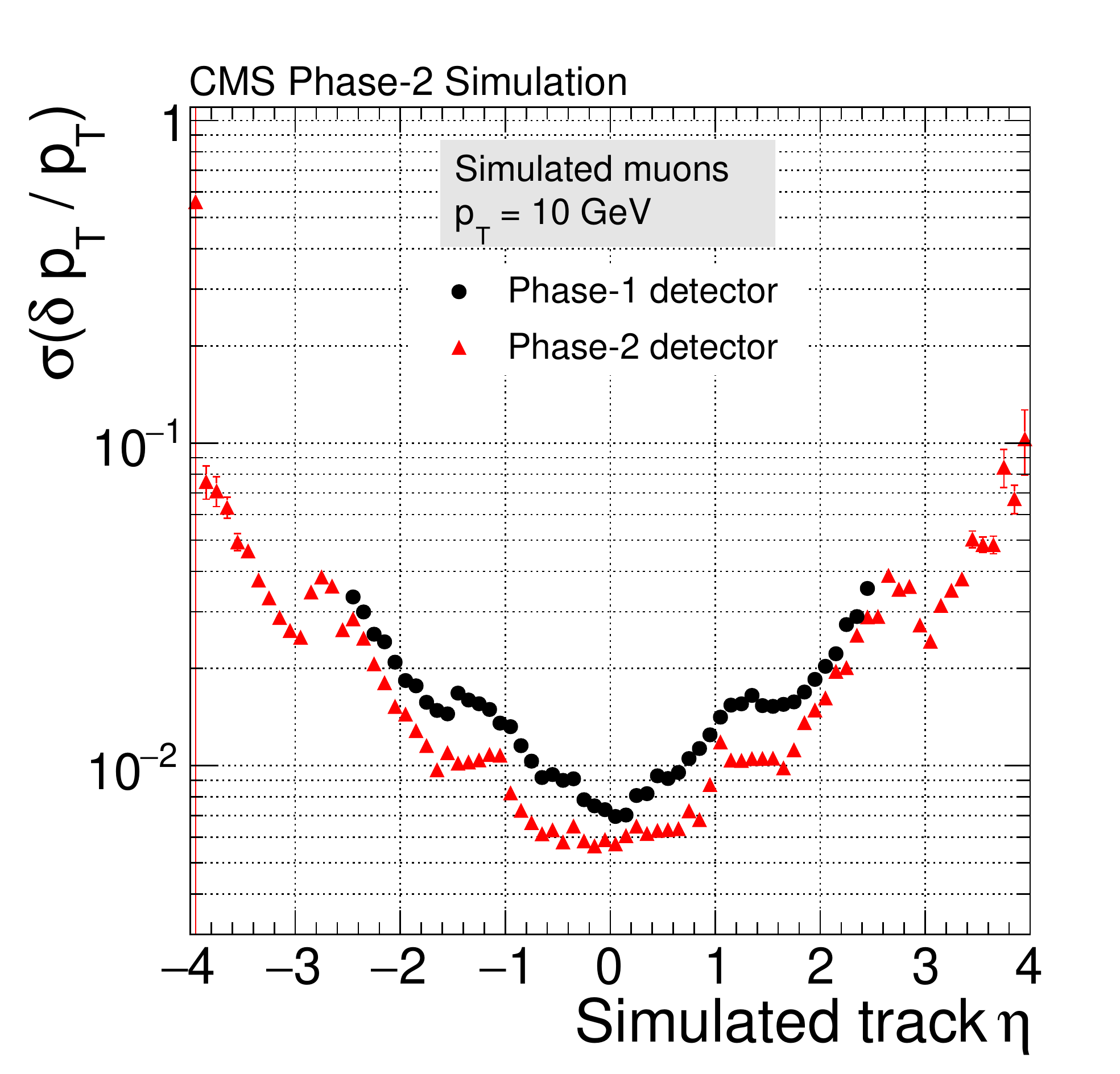}}
  \qquad
  \subfloat[]{\includegraphics[width=0.4\linewidth]{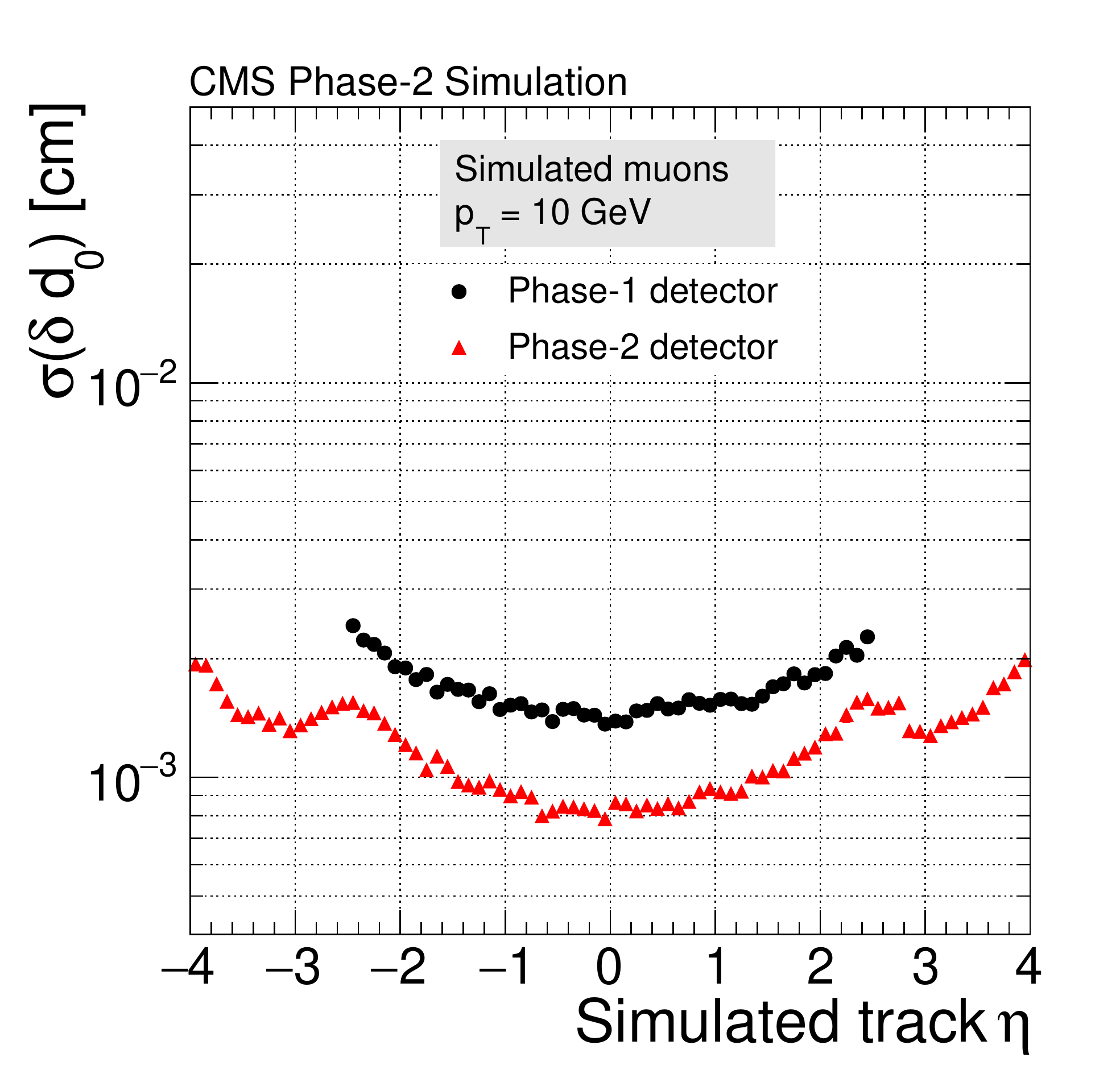}}
  \caption{Comparison of expected resolution on track parameters between the current
    (Phase-1) Tracker and the HL-LHC (Phase-2) Tracker using single isolated
    muons with $p_T$=10~GeV/$c$: transverse momentum (a) and 
    impact parameter in the transverse plane (b).}
  \label{fig:pT_dxy_resolution_singleMuPt10}
\end{figure}

The amount of material in the Tracker volume impacts the accuracy
on the measurement of the track transverse momentum and the rate of fake
tracks and ultimately it is related to the power required to operate the detector,
as high power consumption requires more massive cables for the power distribution  
and a more complex cooling system for efficient heat dissipation.
Despite the estimated power consumption for the upgraded Tracker is
three times larger than for the current Tracker, 50 kW for
the IT and 100 kW for the OT, careful choices for the powering scheme
(serial powering for the IT and use of DC-DC converters for the OT)
and for the cooling system (use of two-phase evaporative CO$_2$ cooling which allows to
use steel cooling pipes with a reduced diameter of about 2~mm) result
in a profile of the material which is even better than for the current
detector. The beneficial effect on the resolution on the track transverse momentum and
on the impact parameter in the transverse plane is shown in
Fig.~\ref{fig:pT_dxy_resolution_singleMuPt10}: the upgraded Tracker
is expected to achieve better performance than the current detector on
an extended pseudo-rapidity range despite the higher pileup expected
at the HL-LHC.

\section{Conclusions}
For the HL-LHC the entire Tracker of the CMS experiment will be
replaced with a new detector which besides having an unprecedented
radiation tolerance will guarantee an efficient online selection and
offline reconstruction of events up to an average pileup of 200
events per bunch crossing.  For the OT the main design choices
have been already established. For the IT crucial decisions like the
choice of the architecture of the front-end chip or the technology of
the sensors will be taken in the next months, mid-2019 and end of 2020 respectively, based on the results from
prototypes that for the first time were operated after being exposed to
radiation levels comparable with those expected at HL-LHC.
In the next years the project will move to the large-scale production
of the components (4k IT modules and 14k OT modules).
Quality assurance during production and care to system aspects in the integration of the different subsystems
will be crucial to guarantee the expected performance of the upgraded Tracker.

\end{document}